# A sentiment analysis of Singapore Presidential Election 2011 using Twitter data with census correction

Murphy Choy[1]
Michelle L.F. Cheong[2]
Ma Nang Laik[3]
Koo Ping Shung[4]

**Abstract**

Sentiment analysis is a new area in text analytics where it focuses on the analysis and understanding of the emotions from the text patterns. This new form of analysis has been widely adopted in customer relation management especially in the context of complaint management. With increasing level of interest in this technology, more and more companies are adopting it and using it to champion their marketing efforts. However, sentiment analysis using twitter has remained extremely difficult to manage due to the sampling bias. In this paper, we will discuss about the application of using reweighting techniques in conjunction with online sentiment divisions to predict the vote percentage that individual candidate will receive. There will be in depth discussion about the various aspects using sentiment analysis to predict outcomes as well as the potential pitfalls in the estimation due to the anonymous nature of the internet.

**Introduction**

Social media has been widely adopted by many private enterprises to market their products as well as services. With the successful campaign of President Obama in the US presidential campaign 2008, social media platforms such as facebook, twitter as well as other sites have catapulted to great success as the leading platform to engage voters. Many political analysts (Stirland, 2008; Pasek, 2006; Xenos, 2007) have attributed his success to the active use of social media to engage voters especially the younger voters whose concerns are usually ignored or given less importance (Pasek, 2006). This combined with the poor adoption of social media by McCain and Palin increased his overall advantage in attracting the more voters (Stirland, 2008). Extreme intricate planning and grassroots activities sealed his campaign success (Stirland, 2008). While originally dismissed as random ranting by youths, twitter has now become the tool of choice for voicing and linking up with people.

Originally, twitter was developed as a micro-blogging tool where users can post a very short blog online to update people who are linked about their status and opinions. All the status updates as well as the opinions written are reflected on the twitter online boards or can be search and extracted using the twitter search API (Twitter, 2011). The twitter search capabilities allow almost real life time searching of the information that is also real time streaming. In the earlier days of twitter, most of the tweets or messages are personal opinions. With the development of the twitter market, a variety of uses such as news and product marketing have proliferated rapidly. Political associations as well as various interest groups have successfully used it to voice their opinions, political positions as well as gathering supports from the online audiences. While there are many analysts who believe that twitter is not very useful (Pearanalytics, 2009), others have lauded the immense potential of twitter (Skemp ,2009).

---

[1] Instructor, School of Information System, Singapore Management University
[2] Practice Associate Professor, School of Information System, Singapore Management University
[3] Practice Assistant Professor, School of Information System, Singapore Management University
[4] Manager, School of Information System, Singapore Management University

There are three major objectives to be achieved using twitter information. The first aim is to assess the amount of information with regards to political election events in a conservative yet well connected country. The second aim is to develop the methodology to reconcile the online information with the political events that transpire to examine how well is the information reflected. The last objective is to use the information gained to predict the incumbent president.

**Background to the Singapore Presidential Election 2011**

In this study, we have collected 16,616 tweets from the twitter which spanned across the period of 17 August till 25 August consisting of the nomination period as well as the campaigning period. This is the second Presidential election since the first one that is held in 1993. After being the President for 12 years, President S.R.Nathan had decided to step due to his age. This resulted in several individuals from various segments of the society coming forth to compete for the position. The President of Singapore is the head of the state and ceremonial in nature. While the President holds several veto powers, the role has limited executive rights and privileges. After the screening process by the Presidential Elections Committee, 4 individuals were identified to be eligible candidates from the applicants and all 4 took part in the Presidential election.

The 4 candidates are Dr Tan Cheng Bock, Dr Tony Tan Keng Yam, Tan Jee Say and Tan Kin Lian. Dr Tan Cheng Bock is a doctor who used to be a member of the parliament for a period of 26 years. He has served in various roles ranging from managing director of private companies to advisor for counseling centers. Dr Tony Tan is a former Deputy Prime Minister of Singapore with experience in political appointments ranging from Education to National Defense. He has served as Deputy Chairman of GIC Singapore which is the sovereign wealth fund of the Nation. Mr Tan Jee Say is a former civil servant turned investment banker. He has served as the Principal Private Secretary of the then Deputy Prime Minister Goh Chok Tong (Former Prime Minister of Singapore) as well as in the administrative arms of the civil service. Mr Tan Kin Lian is the former CEO of NTUC Income Insurance. He has held various directorships at different companies and is a member of various professional bodies. Dr Tony Tan, Dr Tan Cheng Bock and Mr Tan Kin Lian were all former members of the People Action Party while Mr Tan Jee Say is a former member of Singapore Democratic Party.

The Presidential Election is taking place some 3 months since the General Elections 2011. The General Elections 2011 is widely viewed as a historic moment in the political history of Singapore. It was the first time that all the wards except one (Tanjong Pagar GRC) were challenged by opposition parties. A record number of political parties are involved in the election process. It was also the first time that social media was considered as a legal mean of political campaigning and thus fall under the legislation and regulation of the Election rules. This is also the first time that a cooling-day approach was implemented. The incumbent party, PAP, managed to retain its strong majority although there was a negative swing of 7%. It marked the second time that a high ranking and important minister was defeated in the history of Singapore. Social Media was noted to have certain level of influence over the campaign period where several incidents on the Social Media were widely criticised and highlighted the ferocious nature of cyber-political campaigning (Economists, 2011; Scoop, 2011; Fong, 2011). It also highlighted how some online channels are influencing the people (Economists, 2011). Candidates while being non-partisan are still linked by social media to their political affliations.

**Literature Review**

The growth of twitter has increased its overall interest to researchers from disciplines such as sociology, marketing and computer science. There have been multitudes of publications in this area notably in marketing. Among the different research groups, there is a group of researchers who study the effect of social media on the market (Honeycutt and Herring 2009; Nielsen Media Research, 2009). Research has shown that there is a huge variation in the intensity and usage of twitter. The uses of twitters ranged from conversations (Honeycutt and Herring 2009) to word of mouth marketing (Jansen et al. 2009). The main theme of the researches done so far focuses on the generic nature of twitter operating in a function but not specifically specialized to evaluate political themes (Tumasjan et. Al., 2010).

There are wide spread discussions and research about web forum, blogs and twitters as alternative form of political debate. Some researchers have acknowledged the quality of the more prominent political blogs (Woodley, 2008) while others doubted the capabilities of the blogs to aggregate and convey the information (Sunstein, 2008). Research has also shown that while there are active participation in many of the political discussion forum(Fong, 2011) as well as blogs (Jansen and Koop 2009), the population actively participating is very small. At the same time, there was no additional information about the overall relevance of twitter in this case (Tumasjan et. Al., 2010).

Most of the current literature are focused on the the effect of social media on the actual population for issues such as politics, public policies and causes. The literature covered acknowledged the lack of recognition for the non-online population influence on the political landscape (Drezner and Farrell, 2008). Several case studies have found that the online information has been quite successful acting as indicator for electoral success. (Williams and Gulati, 2008). There is very little research on the use of twitter for this purpose (Tumasjan et. Al., 2010).

Therefore the goal of this paper is to:
- Assess the amount of information with regards to political election events in a conservative yet well connected country.
- Develop the methodology to reconcile the online information with the political events that transpire to examine how well is the information reflected.
- Use the information gained to predict the incumbent president.

**Data and Methodology**

We collected twitter information from the start of the campaign on the 17th August 2011 to 25th August 2011. The information was gathered from the twitter search engine with the help of the google API. The data collected is based on the Candidates' name (only the English name is used; this is due to the difficulties of assessing the Chinese tweets.) with a total of 16,616 tweets collected. Repetitive tweeting by unique users are purged and further de-duplications between the different searches are done. This is to ensure that a proper and unadulterated collection of tweets can be used for analysis.

To extract the sentiments from the data automatically, a customized corpus was created and developed for this analysis. While there are several corpus and programs online to conduct sentiment analysis, most of these are not suitable for analyzing this context. Part of the issue with the analysis is the particularly complex abbreviations that are peculiar to this campaign. The localized version of English is slightly different from US or UK English and certain words are used differently. Another problem is the massive use of Parodies which also affects the standard corpus used in the analysis of text information. To this end, a new corpus assembled from online sources, dictionary as well as earlier general elections 2011 campaign was used. It is to be noted that this corpus is strictly for this campaign and not applicable to other campaigns. In this analysis, we will also focus on the sentiment in each tweet and ignore some of the more complicated aspects of tweeting (Tumasjan et. Al., 2010). Due to the possible bias in the data, additional information in the form of census (DOS, 2010) as well as survey from government bodies(IDA, 2010) are collected to correct the inherent bias in the online data.

To estimate the votes, we have developed the following census recorrection framework. In this framework, there are several key information that are required. At the same time, there are 2 assumptions about the framework.

Assumptions
1. The people who voted in the general elections most likely to be voting along the party lines.
2. The online sentiment is representative of the people who are expressing their views.

Both assumptions are due to necessities. If we do not assume that people will vote along their party lines, we will be suggesting an amount of swing voters which we cannot estimate accurately. While polls from both online or offline situation might give some information, it is however uncertain and could yield potentially huge amount of margin error. The second assumption assumes that emotion expressed online represents the viewpoint of the individual and similarly to the people on the ground. This assumption is necessary to avoid further complication in the estimation of the sentiment.

To calculate the percentage of votes that each candidate should receive, we will be using the formula below.

$Let\ X_i =\ total\ population\ in\ age\ group\ i$
$Let\ C_i =\ percentage\ of\ people\ using\ computer\ in\ age\ group\ i$
$Let\ S_i =\ percentage\ of\ people\ using\ social\ media\ in\ age\ group\ i$
$Let\ P_{ni} =\ percentage\ of\ people\ for\ party\ in\ age\ group\ i$
$Let\ C_{ni} =\ percentage\ of\ people\ for\ candidate\ C\ who\ are\ from\ party\ in\ age\ group\ i$
$Let\ E_c =\ percentage\ of\ support\ from\ twitter\ for\ candidate\ C\ within\ party$
$Let\ T_c =\ percentage\ of\ people\ for\ candidate\ C$

From the above, we obtain the constituent components as below.

$Twitter\ Online\ Support\ =\ P_{ni}C_iS_iE_c$ \hfill (1)
$Non\ Twitter\ Online\ Support\ =\ P_{ni}C_i(1-S_i)E_c$ \hfill (2)
$Offline\ Support\ =\ P_{ni}(1-C_i)E_c$ \hfill (3)

Combining equation (1), (2) and (3),

$$C_{pni} = \{P_{ni}C_iS_iE_c + P_{ni}C_i(1-S_i)E_c + P_{ni}(1-C_i)E_c\} \quad (4)$$

Consolidating the information from each age group, we obtain

$$T_c = \sum_{1}^{i}\sum_{1}^{n} C_{ni} \quad (5)$$

We will use the above framework to model the Presidential Election 2011.

**Result**

In this section, we will evaluate all the information made available and use the framework to calculate the estimated percentage of vote for each candidate.

Based on the preliminary information from the census and surveys, table 1 below shows the population percentages as well as the computer literacy rate.

| Age Group (Years) | Total ('000) | % Pop | Comp Literacy (IDA, 2009) |
|---|---|---|---|
| 0 - 4 | 194.4 | 5.15% | 99% |
| 5 - 9 | 215.7 | 5.72% | 99% |

| Age Group | Population (thousands) | % of Total | Computer Literacy |
|---|---|---|---|
| 10 - 14 | 244.3 | 6.48% | 99% |
| 15 - 19 | 263.8 | 6.99% | 97% |
| 20 - 24 | 247.2 | 6.55% | 97% |
| 25 - 29 | 272.6 | 7.23% | 95% |
| 30 - 34 | 298.7 | 7.92% | 95% |
| 35 - 39 | 320.0 | 8.48% | 76% |
| 40 - 44 | 309.4 | 8.20% | 76% |
| 45 - 49 | 323.5 | 8.58% | 76% |
| 50 - 54 | 303.0 | 8.03% | 44% |
| 55 - 59 | 248.7 | 6.59% | 44% |
| 60 - 64 | 192.0 | 5.09% | 14% |
| 65 - 69 | 111.5 | 2.96% | 14% |
| 70 - 74 | 92.6 | 2.46% | 14% |
| 75 - 79 | 65.2 | 1.73% | 14% |
| 80 - 84 | 39.8 | 1.06% | 14% |
| 85 & Over | 29.2 | 0.77% | 14% |
| Total | 3,771.7 | | |

Table 1: Population percentages and computer literacy rate

However, due to the secrecy of the voting process and the obvious lack of polls, we were not able to obtain direct information of the popular support for the parties. As a result of this, we use some of the information obtained from the internet about the voting populace. One of the key information is that the older age groups are the more staunch supporters of PAP. At the same time, the general elections 2011 also indicated that the younger generations are less supportive with the general support for PAP to be at 60.1%. Using this information, we use a simple linear programming form to calculate the support for PAP.

Let $P_i$ = proportion of PAP supporter in age group $i$
Let $X_i$ = Proportion of population in age group $i$

$$Objective: \frac{\sum_1^i X_i P_i}{\sum_1^i X_i} = 0.6$$

Constraints:

$\forall\, i, j \in i > j, P_i > P_j$
$\forall\, i, 0 < P_i < 1$

Using this optimization process, we obtain the following result of the PAP supporters in table 2. While these are the numbers we derived, there are several possible solutions to this problem.

| Age Group | Est. % of PAP Supporter | % of Pop Online | % Non Social Media | % Social Media |
|---|---|---|---|---|

| (Years) | | | | |
|---|---|---|---|---|
| 20 - 24 | 43.3% | 6.4% | 2.9% | 3.4% |
| 25 - 29 | 46.0% | 6.9% | 4.5% | 2.4% |
| 30 - 34 | 48.8% | 7.5% | 4.9% | 2.6% |
| 35 - 39 | 51.9% | 6.4% | 5.4% | 1.1% |
| 40 - 44 | 55.0% | 6.2% | 5.2% | 1.1% |
| 45 - 49 | 58.4% | 6.5% | 5.4% | 1.1% |
| 50 - 54 | 62.0% | 3.5% | 3.3% | 0.2% |
| 55 - 59 | 65.8% | 2.9% | 2.7% | 0.2% |
| 60 - 64 | 69.9% | 0.7% | 0.7% | 0.0% |
| 65 - 69 | 74.2% | 0.4% | 0.4% | 0.0% |
| 70 - 74 | 78.8% | 0.3% | 0.3% | 0.0% |
| 75 - 79 | 83.6% | 0.2% | 0.2% | 0.0% |
| 80 - 84 | 88.7% | 0.1% | 0.1% | 0.0% |
| 85 & Over | 94.2% | 0.1% | 0.1% | 0.0% |
| Overall % | 60.0% | | | |

Table 2: Table showing the estimated percentage of PAP supporters as well as the breakdown of online, social media active and inactive percentages.

The population online is then calculated in this case using the computer literacy percentages. Active social media population and non active population are further separated using the information from surveys (IDA, 2009). For the off tweet population, we will separate them into the PAP and non-PAP supporter by the same ratio as the general elections 2011 as shown in table 3.

| Age Group (Years) | PAP % | Opp % |
|---|---|---|
| 20 - 24 | 1.4% | 1.8% |
| 25 - 29 | 2.2% | 2.6% |
| 30 - 34 | 2.6% | 2.7% |
| 35 - 39 | 3.8% | 3.6% |
| 40 - 44 | 3.9% | 3.2% |
| 45 - 49 | 4.4% | 3.1% |
| 50 - 54 | 4.8% | 3.0% |
| 55 - 59 | 4.2% | 2.2% |
| 60 - 64 | 3.5% | 1.5% |
| 65 - 69 | 2.2% | 0.8% |
| 70 - 74 | 1.9% | 0.5% |
| 75 - 79 | 1.4% | 0.3% |
| 80 - 84 | 0.9% | 0.1% |
| 85 & Over | 0.7% | 0.0% |
| **Total Pop %** | **38.0%** | **25.3%** |

Table 3: Offline population spread between party lines.

Using the online sentiments, we calculate the overall sentiment for each candidate within their respective party line.

| Grouping | Candidate | Sentiment (Value) | % Split |
|---|---|---|---|
| 1 | TT | 275 | 49.1% |
| 1 | TCB | 285 | 50.9% |
| 2 | TJS | 356 | 59.3% |
| 2 | TKL | 244 | 40.7% |

Table 4: Candidates' online twitter positive sentiments

There are several points to note about the sentiment value. The sentiment value is the aggregated positive and negative emotion from the tweets. One or multiple candidates may be endorsed in a tweet while there could be mixed endorsements. Correlation analysis between the candidates also indicated that there are no strong correlations between the tweets of the candidates. Tweets which contain endorsement for all candidates were removed as they provide no critical useful information.

Given the splits in table 4, we then apply the information into framework.

|  | Offline | | | | Online | | | | Overall | | | |
|---|---|---|---|---|---|---|---|---|---|---|---|---|
| Age Group (Years) | TT | TCB | TJS | TKL | TT | TCB | TJS | TKL | TT | TCB | TJS | TKL |
| 20 - 24 | 0.66% | 0.69% | 1.04% | 0.72% | 0.73% | 0.76% | 1.15% | 0.79% | 1.39% | 1.45% | 2.20% | 1.52% |
| 25 - 29 | 1.09% | 1.13% | 1.54% | 1.07% | 0.54% | 0.56% | 0.77% | 0.53% | 1.63% | 1.69% | 2.31% | 1.60% |
| 30 - 34 | 1.27% | 1.32% | 1.60% | 1.11% | 0.63% | 0.65% | 0.80% | 0.55% | 1.90% | 1.97% | 2.39% | 1.66% |
| 35 - 39 | 1.88% | 1.95% | 2.10% | 1.46% | 0.28% | 0.29% | 0.31% | 0.21% | 2.16% | 2.24% | 2.41% | 1.67% |
| 40 - 44 | 1.93% | 2.01% | 1.89% | 1.32% | 0.29% | 0.30% | 0.28% | 0.19% | 2.21% | 2.30% | 2.18% | 1.51% |
| 45 - 49 | 2.14% | 2.23% | 1.83% | 1.27% | 0.32% | 0.33% | 0.27% | 0.19% | 2.46% | 2.56% | 2.11% | 1.46% |
| 50 - 54 | 2.37% | 2.46% | 1.74% | 1.21% | 0.08% | 0.08% | 0.06% | 0.04% | 2.44% | 2.54% | 1.80% | 1.25% |
| 55 - 59 | 2.06% | 2.15% | 1.29% | 0.90% | 0.07% | 0.07% | 0.04% | 0.03% | 2.13% | 2.21% | 1.33% | 0.92% |
| 60 - 64 | 1.73% | 1.80% | 0.90% | 0.62% | 0.01% | 0.02% | 0.01% | 0.01% | 1.74% | 1.81% | 0.90% | 0.63% |
| 65 - 69 | 1.07% | 1.11% | 0.45% | 0.31% | 0.01% | 0.01% | 0.00% | 0.00% | 1.07% | 1.12% | 0.45% | 0.31% |
| 70 - 74 | 0.94% | 0.98% | 0.31% | 0.21% | 0.01% | 0.01% | 0.00% | 0.00% | 0.95% | 0.99% | 0.31% | 0.21% |
| 75 - 79 | 0.70% | 0.73% | 0.17% | 0.12% | 0.01% | 0.01% | 0.00% | 0.00% | 0.71% | 0.74% | 0.17% | 0.12% |
| 80 - 84 | 0.46% | 0.47% | 0.07% | 0.05% | 0.00% | 0.00% | 0.00% | 0.00% | 0.46% | 0.48% | 0.07% | 0.05% |
| 85 & Over | 0.35% | 0.37% | 0.03% | 0.02% | 0.00% | 0.00% | 0.00% | 0.00% | 0.36% | 0.37% | 0.03% | 0.02% |
| Total Pop % | | | | | | | | | 21.61% | 22.47% | 18.65% | 12.92% |
| Election % | | | | | | | | | 28.6% | 29.7% | 24.7% | 17.1% |

Table 5: Estimated results

Using the model predicted, we then validated it against the final result which is released on 28 August 2011 morning as shown in table 6.

|     | Predicted | Actual |
| --- | --- | --- |
| TT  | 28.60% | 35.19% |
| TCB | 29.70% | 34.85% |
| TJS | 24.70% | 25.04% |
| TKL | 17.10% | 4.91% |

Table 6: Result comparison

From the results, we can see the difference between the predicted and the actual value. While we have successfully calculated the thin margin between the top two candidates, the model did not succeed in predicting the right winner. At the same time, one of the candidates also has much lower results than the predicted result. We will offer some explanations to the discrepancies.

The first problem which the model is currently unable to address is the percentage of swing voters. This group of voters does not vote according to the party line and often change their position depending on the policies of the parties. At the same time, due to the non-partisan nature of the presidency, some of the swing voters might not have voted as they have done previously. This could account for the differences in the actual vs. predicted.

The second problem is the issue with fake tweeter sentiment. This situation can be attributed to 2 major causes. The first source of such fake sentiment is astroturfing. The other source may be related to the scenario where the voters do not truly reflect their online sentiments from their choice of candidate. This seems to be a peculiar situation that does not occur elsewhere.

**Conclusion**

From the results, the framework has been able to predict the top two contenders in the four corner fight. While the predicted winner does not emerge as the president due the small margin, the estimation of the small margin of votes between the contenders indicated the model's ability to realistically model the scenario. The framework has been able to convert the twitter information into realistic prediction. At the same time, this is the first time that twitter information has been collected in a conservative highly connected environment where a significant portion of the population are not involved in the twitter movement. The analysis has proved that given proper recalibration using census information, the twitter information can translate into pretty accurate information about the political landscape even though the twitter users are not as common. Further work will be needed in order to modify the model slightly in order to calibrate the swing voters into the equation. At the same time, more important work will be needed to handle the issue of astroturfing as well the psychology of voters.